\documentclass[reprint,
superscriptaddress,
 amsmath,amssymb,
 aps,
 prl,
]{revtex4-2}

\usepackage{xcolor}
\usepackage{graphicx}
\usepackage{dcolumn}
\usepackage{bm}

\begin{document}


\title{Generalizing multiple memories from a single drive: The hysteron latch}

\author{Chloe W.\ Lindeman}
\email{cwlindeman@uchicago.edu}
\affiliation{James Franck Institute and Department of Physics, University of Chicago, Chicago IL, USA}
\author{Travis R.\ Jalowiec}
\email{tjalowiec@nanohmics.com}
\altaffiliation{Nanohmics, Inc., Austin TX, USA}
\affiliation{Department of Physics, Pennsylvania State University, University Park, PA, USA}
\author{Nathan C.\ Keim}
\email{keim@psu.edu}
\affiliation{Department of Physics, Pennsylvania State University, University Park, PA, USA}

\date{\today}

\begin{abstract}

Far-from-equilibrium systems can form memories of previous deformations or driving. In  systems from sheared glassy materials to buckling beams to crumpled sheets, this behavior is dominated by return-point memory, in which revisiting a past extremum of driving restores the system to a previous state. 
Cyclic driving with both positive and negative strains forms multiple nested memories---as in a single-dial combination lock---while asymmetric driving (only positive strain) cannot. 
We study this case in a general model of hysteresis that considers discrete elements called hysterons. 
We show how two hysterons with a frustrated interaction can violate return-point memory, realizing multiple memories of asymmetric driving. This reveals a general principle for designing systems that store sequences of cyclic driving, whether symmetric or asymmetric. In disordered systems, asymmetric driving is a sensitive tool for the direct measurement of frustration.

\end{abstract}

\maketitle

\begin{itemize}
    \item[Short title:] Generalizing multiple memories from a single drive
    \item[Teaser:] Adding frustrated interactions within a system with hysteresis reveals a general principle for storing multiple values from the history of a single variable, as in a combination lock.
\end{itemize}

\section{Introduction}

The single-dial combination lock is a mechanism for storing multiple values from a single input. By alternating between clockwise and counterclockwise rotation, the operator stores the combination values as a series of turning points. Each new turning point must be nested within the previous two, so that the lock verifies not only the values but their exact sequence. Information about the stored values can be recovered by observing the resistance to further rotation~\cite{Keim2019MemoryMatter}. Finally, erasure can be achieved with a large twist of the dial. Every operation is accomplished with a single control.

This elegant idea was known by 1909~\cite{Junkunc1909}, but it was rediscovered decades later as return-point memory: a general principle for systems of hysteretic elements driven by a field that alternately increases and decreases~\cite{barker1983,Sethna1993HysteresisTransformations,Keim2019MemoryMatter}. Applications have ranged from ferromagnets~\cite{barker1983,Preisach1935} to rocks~\cite{Guyer2006}, and more recently, memories of deformation in amorphous solids~\cite{Keim2020GlobalSolid,Mungan2019NetworksRemember}, crumpled sheets~\cite{Shohat2022MemorySheets}, and designed metamaterials that feature origami folds or buckling beams~\cite{bense2021complex,Jules2022Origami}.
In these newer examples, observations tied each system’s memory of mechanical strain to the hysteretic elements within it: particle rearrangements, creases, or buckling units, all modeled as bistable ``hysterons.’’

\begin{figure}
    \begin{center}
    \includegraphics[width=3.4in]{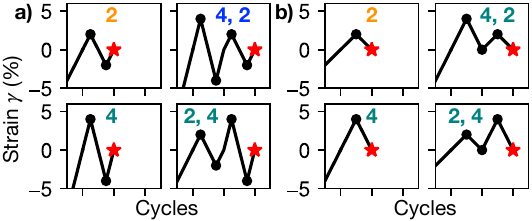}
    \end{center}
    \caption{\textbf{Driving protocols can store multiple memories.}
    The state of a generic system with return-point memory depends on the sequence of nested turning points from driving.
\textbf{(a)} Driving protocols that store the amplitude(s) of symmetric (positive and negative) shear; dots mark turning points. Final memory-encoding state corresponds to $\bigstar$. ``2’’, ``4’’, and ``4, 2’’ each lead the system to a different state. ``2, 4'' does not nest turning points within preceding ones, and so yields the same state as ``4''.
\textbf{(b)} Asymmetric equivalents of (a). Because every cycle has the same turning point at 0, it is impossible to store multiple values via return-point memory.
    \label{fig:protocol}}
\end{figure}

\begin{figure}
    \begin{center}
    \includegraphics[width=3.4in]{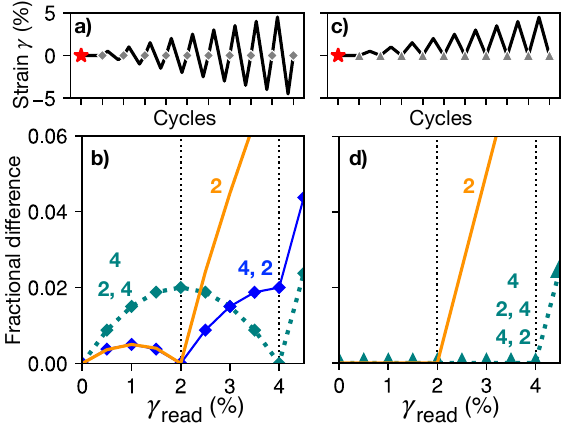}
    \end{center}
    \caption{\textbf{Readout confirms the turning-point analysis of Fig.~\ref{fig:protocol}.}
    \textbf{(a)} Symmetric strain protocol for reading out memories. State $S_\text{mem}$ after writing, marked with $\bigstar$, is compared with state after each cycle (gray diamonds), as amplitude $\gamma_\text{read}$ increases.
    \textbf{(b)} Results of readout, measured as fraction of hysterons that differ. Each curve is labeled with the write protocol(s) from Fig.~\ref{fig:protocol} that produced it.
    The most recent amplitude is always present, and multiple memories are possible if written in descending order.
    \textbf{(c)} Equivalent asymmetric protocol. Ends of readout cycles are marked with gray triangles.
    \textbf{(d)} Readout results for each protocol in (c). The system cannot store a sequence of asymmetric driving, and the most recent amplitude may be missing. 
    \label{fig:preisach}}
\end{figure}

In this paper, we consider a simple change to the driving that leaves the combination lock or any other system with return-point memory unable to store multiple values. In conventional \emph{symmetric} driving values are introduced by changing both the upper and lower turning points of motion (Fig.~\ref{fig:protocol}a). In \emph{asymmetric} driving one turning point is kept fixed (Fig.~\ref{fig:protocol}b), such as when a bridge is crossed by a series of vehicles with different weights. This simple change allows no memories but the largest one, a fact we verify by writing and reading memories in ensembles of model hysterons~\cite{Keim2011GenericNoise,Adhikari2018MemoryAssemblies,Keim2020GlobalSolid} (Fig.~\ref{fig:preisach}). 
However, in place of return-point memory arising from single hysterons, for asymmetric driving we find an equivalent mechanism based on coupled \emph{pairs} of hysterons, allowing a system to store multiple driving amplitudes and ensuring that it always remembers the most recent input. We then formulate a general ``latching’’ principle for storing multiple memories sequentially from a single drive, whether symmetric or asymmetric. Our results point the way to new studies of glassy physics in disordered matter, and to the rational design of new information-processing capabilities in mechanical systems.


\section{Results}

We are motivated by models of disordered solid materials and metamaterials, wherein a system contains mesoscopic ``hysterons'' that can
reversibly switch between two states under an imposed strain $\gamma$---for example by rearrangements of particles in an amorphous solid~\cite{nicolas2018,Mungan2019NetworksRemember,khirallah2021}, or snap-through of a buckled beam~\cite{bense2021complex,Merrigan2021Metamaterial}. These models capture many mesoscopic details of experiments and simulations~\cite{Mungan2019NetworksRemember,Keim2020GlobalSolid,Regev2021Topology,Szulc2022CooperativeSolids,Jules2022Origami,bense2021complex,Liu2024}, especially when hysterons can interact. The $i$th hysteron switches its state $S_i$ at thresholds $\gamma_i^\pm$: when $\gamma > \gamma_i^+$, $S_i = +1$; when $\gamma < \gamma_i^-$, $S_i = -1$; and when $\gamma_i^- < \gamma < \gamma_i^+$, $S_i$ remains in its previous state. 
For convenience we write states as ``$+$'' and ``$-$''. 
In our simulations, the ``system'' is an ensemble of many groups of $N$ hysterons with random parameters, modeling interactions via perturbed thresholds

\begin{equation}
\gamma_i^\pm(S_{j \neq i}) = \gamma^\pm_i - \sum_{j \neq i} J_{ij} S_j
\end{equation}

\noindent where $S_{j \neq i}$ represents the states of all hysterons excluding hysteron $i$, and $J_{ij}$ is the $N\times N$ interaction matrix. This is equivalent to the model in Ref.~\cite{Keim2021MultiperiodicSolids}. $J_{ij} < 0$ represents a frustrated interaction, where one hysteron flip inhibits another from flipping. We choose $J_{ij}$ with uniform probability from $[-J_0, 0]$, with $J_0 = 0.01$ unless otherwise specified.
Each hysteron's $\gamma^\pm_i$ are chosen by drawing two values from a uniform distribution on $[-0.1, 0.1]$, and then ordering them so that $\gamma_i^-$ $<$ $\gamma_i^+$, corresponding to rearrangements that dissipate energy. 

\subsection{Memories in Non-Interacting and Interacting Systems}

Protocols for initializing each simulation and writing memories are shown in Fig.~\ref{fig:protocol}; example readout protocols are in Fig.~\ref{fig:preisach}(a, c). We begin each simulation with all hysterons in the ``$-$'' state, as though $\gamma \to -\infty$, and then we drive the hysterons to $\gamma = 0$ before encoding memories. (Starting at $\gamma \to \infty$ similarly supports our conclusions; it is considered in the Supplemental Materials~\cite{SM}.)
Simulations used the open-source \texttt{hysteron} software package~\cite{Keim2021MultiperiodicSolids}.
Like molecular dynamics simulations of amorphous solids, we work in the athermal and quasistatic limit: the algorithm identifies the hysteron that will flip soonest as $\gamma$ changes, and then holds $\gamma$ fixed while it updates any other hysterons that were destabilized via interactions (e.g.\ an avalanche), starting with the hysteron farthest past its threshold.
When $N > 2$, the random parameters would occasionally prevent the algorithm from finding a stable state; these cases were discarded~\cite{Keim2021MultiperiodicSolids}. 
We did not consider cases where $J_{ij}$ and $J_{ji}$ have opposite signs, for which this issue is common.  

Without loss of generality, we take the single stored strain amplitude to be 4\%, and in two-memory tests we use 4\% and 2\%. After the writing cycle(s), the state is saved as $S_\text{mem}$. To read out the memories encoded in $S_\text{mem}$, we apply a series of cycles with increasing amplitude $\gamma_\text{read} = 0,\ 0.005,\ 0.01\dots$. This is a ``serial'' protocol that is suited to experiments, as opposed to a ``parallel'' protocol in which a separate copy of the system is made for each readout cycle~\cite{Adhikari2018MemoryAssemblies}. 
After each cycle with amplitude $\gamma_\text{read}$, we record the fraction of hysterons that are different from $S_\text{mem}$. We average this fraction over the entire ensemble and plot it against $\gamma_\text{read}$.

\begin{figure}
\includegraphics[width=3.2in]{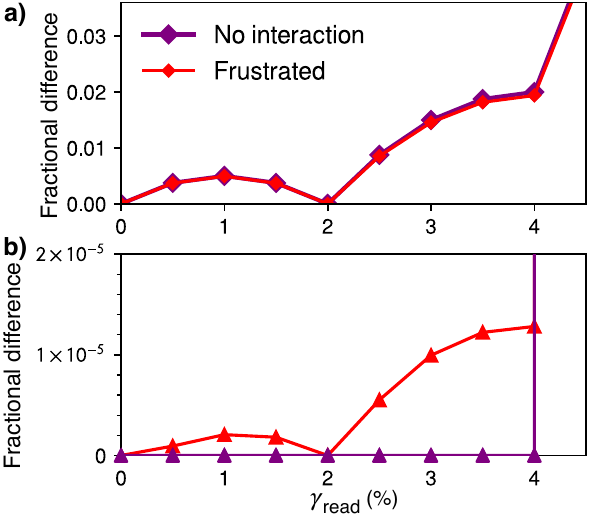}
\caption{
\textbf{Frustrated hysteron pairs store multiple amplitudes of both symmetric and asymmetric driving.}
\textbf{(a)} Curves from symmetric driving have cusps at 2\% and 4\%.
\textbf{(b)} Asymmetric driving. Without interactions, only the memory at 4\% is present. Curve from $10^8$ frustrated pairs shows both memories.
}
\label{fig:double}
\end{figure}

Figure~\ref{fig:double}a shows readouts after the symmetric two-amplitude protocol labeled ``4, 2'' in Fig.~\ref{fig:protocol}a. The curve from an ensemble of non-interacting hysterons ($J_{ij} = 0$) is reproduced from the ``4, 2'' curve of Fig.~\ref{fig:preisach}b; it features a local minimum that indicates the most recent memory ($2\%$) and a cusp that indicates the larger memory ($4\%$), just as in the corresponding amorphous solid experiments and molecular dynamics simulations~\cite{Fiocco2014EncodingSolids,Adhikari2018MemoryAssemblies,Keim2020GlobalSolid,Keim2022MechanicalSolid}. However, in Fig.~\ref{fig:double}b driving the same ensemble asymmetrically changes its behavior dramatically: the curve shows no signature of the smaller driving amplitude, rising rapidly only after $4\%$. 

These strikingly different behaviors are both expected from return-point memory, wherein the system remembers the turning points of driving. As long as $\gamma$ is bounded between any pair of turning points, visiting either turning point will return the system to the state it had when it was at that turning point before~\cite{barker1983,Sethna1993HysteresisTransformations}. This property is recursive, meaning that one may encode more than one memory with a \emph{symmetric} driving protocol by decreasing the strain amplitude, such that each new pair of turning points is nested within the last as in Fig.~\ref{fig:preisach}a. 

However, return-point memory also means that two \emph{asymmetric} cycles can write only one memory: the first cycle of the ``4, 2'' protocol in Fig.~\ref{fig:protocol}b establishes a bounding turning point at $\gamma=0$, and visiting $\gamma=0$ again after writing the second, smaller amplitude immediately restores the state with just one memory. Repeating the results of Fig.~\ref{fig:preisach}d, the readout of non-interacting hysterons in Fig.~\ref{fig:double}b fails to change their states until $\gamma=4\%$ is exceeded---writing a second, smaller memory has no effect. 

We now consider frustrated interactions, $J_{ij} < 0$, as found in models of ``glassy'' matter such as crumpled sheets, disordered or amorphous solids, or spin ice and spin glass. 
Frustration means that one relaxation inhibits others, leading to these materials' characteristically rugged landscapes of metastable states with broad distributions of energy barriers~\cite{binder1986}. 
In hysteron models, the sequences in which hysterons switch during forward or reverse shear become mutable, so that return-point memory is no longer assured~\cite{Sethna1993HysteresisTransformations,hovorka2005,vanHecke2021ProfusionHysterons,Szulc2022CooperativeSolids}. 
The red curves of Fig.~\ref{fig:double}a show that nonetheless, replacing the ensemble of single hysterons with an ensemble of frustrated pairs merely perturbs the return-point memory of symmetric driving. However, in Fig.~\ref{fig:double}b where the readout curve for asymmetric driving had been zero, there is now a clear signature of both memories. This new signal resembles the much larger one from symmetric driving, suggesting a connection.

\begin{figure*}
    \includegraphics[width=7in]{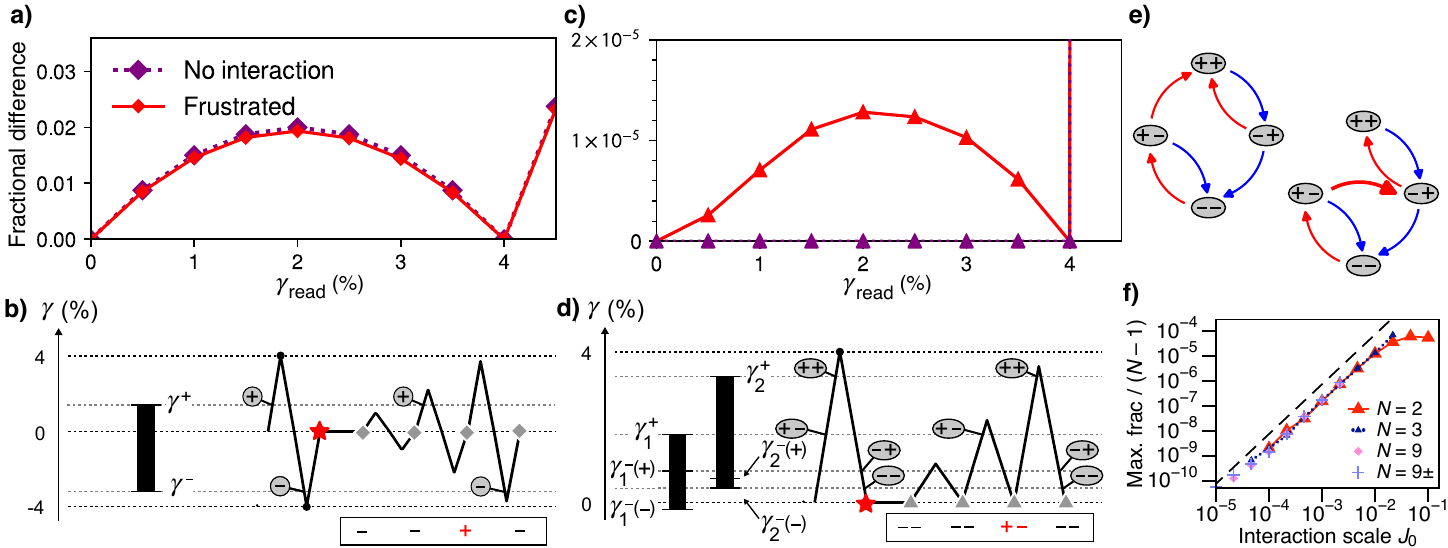} 
    \caption{
    \textbf{Mechanisms for memory.}
    \textbf{(a)} Readouts of single memory of symmetric driving. Local minima indicate the memory.
    \textbf{(b)} Schematic of how a single hysteron contributes to memory, showing hysteron thresholds on the left and an abridged driving protocol on the right. Changes in state during readout are marked with circled $+$ and $-$. State at the end of each cycle is shown below; only the $+$ state contributes to readout. 
    \textbf{(c)} Readouts of single memory with asymmetric driving. The curve without interactions is zero for $\gamma_\text{read} \leq 4\%$ and then increases, while the curve with frustration is non-monotonic.
    \textbf{(d)} Schematic for a frustrated pair with asymmetric driving. For clarity, we show only how the interaction splits the hysteron bottoms $\gamma^-_1$ and $\gamma^-_2$ as a function of the other hysteron's state, since the ordering of these four values is crucial for non-zero readout. The hysteron tops $\gamma_1^+$ and $\gamma_2^+$ can vary widely; see~\cite{SM} for details. 
    \textbf{(e)} Transition graphs that can give rise to non-zero readout below the remembered strain in simulations of frustrated hysterons. Transitions while increasing (decreasing) strain are in red (blue). Left graph corresponds to (d). Thicker arrow represents an avalanche in which the intermediate state $-+$ falls immediately into $--$.
    \textbf{(f)} Peak signal in (c) for $\gamma_\text{read} < 4\%$, varying interaction scale $J_0$ and number of mutually-interacting hysterons $N$, scaled by the number of pair interactions per hysteron ($N-1$). In ``$9\pm$,'' half of the $J_{ij}, J_{ji}$ pairs are positive (cooperative), so data are scaled by $(N-1)/2$. Each point is the average of $10^9 N$ hysterons. Dashed line shows $J_0^2$ scaling for reference.
    }
    \label{fig:mech}
\end{figure*}

\subsection{Memory Mechanisms}

To understand this connection mechanistically, we 
first return to non-interacting hysterons and examine their memory of a single amplitude, encoded with the symmetric protocol labeled ``4'' in Fig.~\ref{fig:protocol}a. Just as in the two-memory case, the corresponding readout curve in Fig.~\ref{fig:mech}a is consistent with return-point memory:
each readout cycle with $\gamma_\text{read} < 4\%$ establishes new turning points nested within the original pair at $\pm 4\%$, placing the system in a new and distinct state and yielding a nonzero difference signal; when $\gamma_\text{read} = 4\%$ the original turning points are revisited and the state at the cycle's end matches $S_\text{mem}$, making the signal zero. This leads to the distinctive rise and fall of the readout curve below the training strain.


In this case, we can understand the behavior of the ensemble by studying how a single hysteron contributes to memory.
Figure~\ref{fig:mech}b shows the response of a particular hysteron to an abridged writing and readout protocol. Only hysterons such as this one, with $-\gamma^- > \gamma^+ > 0$, contribute to readout for $\gamma_\text{read} \le 4\%$ because only they will be in the $-$ state after writing, will then become trapped in the $+$ state when amplitude is reduced, and will finally return to $-$ when the original amplitude is resumed. In this sense, the hysteron ``latches'' into the $+$ state during intermediate strain cycles.
From the broad distributions of $\gamma^\pm$, this mechanism yields the smooth rising and falling curve for $\gamma_\text{read} \le 4\%$ in Fig.~\ref{fig:mech}a. The hysterons with other arrangements of $\gamma^\pm$ end each readout cycle in the same state for $0 \le \gamma_\text{read} \le 4\%$ and hence do not contribute to the signal below the training strain.


Just as with two memories, in Fig.~\ref{fig:mech}c frustration enables a single memory of asymmetric driving that resembles its symmetric counterpart. We show in Fig.~\ref{fig:mech}d that this happens by an analogous mechanism involving a frustrated pair of hysterons. Frustration allows a \textit{two-hysteron} latching behavior in which revisiting the turning point at $\gamma=0$ can fail to restore the previous state. 
The mechanisms of Figs.~\ref{fig:mech}b and \ref{fig:mech}c are thus equivalent when one treats each whole cycle as one transition, either to the same state or to a new state~\cite{Paulsen2019MinimalMemories}.

\subsection{Scaling the Two-Hysteron Latch}

What features are needed for latching? 
The values $\gamma^+_1$ and $\gamma^+_2$ set the hysterons' sensitivity to amplitude; as long as they exceed most lower thresholds they can vary widely, creating the smooth, slightly asymmetric curves in Fig.~\ref{fig:mech}c. By contrast, the lower thresholds must satisfy 
\begin{equation}
\gamma^-_1(-)< 0 < \gamma^-_2(-) < \gamma^-_2(+) < \gamma^-_1(+),
\label{eq:simpler_pair}
\end{equation}
as in Fig.~\ref{fig:mech}d. These inequalities were verified in simulations by testing $10^9$ random pairs. In addition to the explicit flipping thresholds shown, it can be useful to visualize the order in which states are visited via a ``transition graph.'' The graphs that can lead to latching in our frustrated simulations---and hence to non-zero readout---are shown in
Fig.~\ref{fig:mech}e. If we let $J_{ij}$ and $J_{ji}$ have opposite signs, a third transition graph can contribute (see Supplementary Materials for details~\cite{SM}). 

The interaction strength sets the threshold ``window'' size $\gamma^-_1(+) - \gamma^-_1(-) = -2J_{12}$, into which both 0 and the interval $[\gamma_2^-(-), \gamma_2^-(+)]$ must fall. For ensembles with uniformly-distributed parameters like those reported here, these two requirements make the probability $P$ for Eq.~\ref{eq:simpler_pair} second-order in the interaction strength, i.e.\ $P \sim J_0^2$. 
The signals in Fig.~\ref{fig:mech}c are thus small compared to the result from return-point memory under symmetric driving, which is zeroth-order in the sense that it may be obtained with $J_0 = 0$. The $J_0^2$ scaling is confirmed for small interaction strength ($J_0 \ll 4\%$) in  Fig.~\ref{fig:mech}f, where we measure the maximum height of the readout signal for $\gamma_\text{read} < 4\%$~\cite{SM}. In a design context, our analysis means that greatest tolerance for manufacturing errors corresponds to large $J_{12}$ and small $J_{21}$, while keeping $\gamma^-_1(+)$ below the smallest amplitude to be remembered. Crucially, hysteron pairs that fail to satisfy Eq.~\ref{eq:simpler_pair} do not corrupt an ensemble's memory; they are simply absent from readout.

The scaling estimate presented above is a departure from analyses of $P$ based solely on the ordering of thresholds, as considered by van Hecke~\cite{vanHecke2021ProfusionHysterons}: here we included the turning point of asymmetric driving, which additionally isolates the $J_0^2$ behavior by cutting out the zeroth-order response. 

Remarkably, the two-hysteron latch is also how memories of asymmetric driving arise in larger groups of interacting hysterons, so that this motif may be observed in a bulk disordered material or metamaterial~\cite{Merrigan2021Metamaterial,Sirote-Katz2024}. Figure \ref{fig:mech}f shows nearly identical results for larger, mutually-interacting groups after dividing out the multiplicity of frustrated pairs. These results hold even when we randomly make half of the interactions cooperative (drawn from $[0, J_0]$)---strongly suggesting that Fig.~\ref{fig:mech}d is the dominant mechanism despite many more possible behaviors~\cite{vanHecke2021ProfusionHysterons,Keim2021MultiperiodicSolids,Lindeman2021MultipleLandscapes,Szulc2022CooperativeSolids}. 
In the Supplemental Materials~\cite{SM} we further show that the memory-forming portions of these larger groups tend to have the same kinematics and interaction strengths as in $N=2$.
This remarkable conservation is possible because hysterons that do \emph{not} contribute to asymmetric readout are largely following return-point memory, and so their states and transitions vary little from cycle to cycle.

\begin{figure}
\includegraphics[width=3.2in]{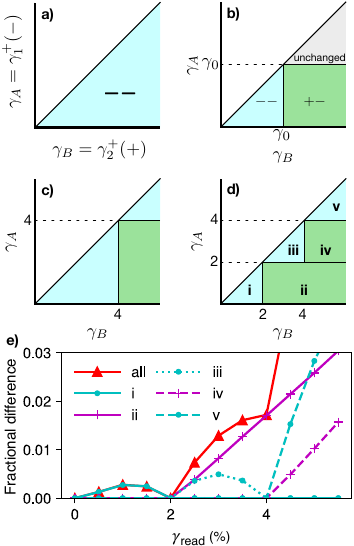}
\caption{\textbf{Graphical analysis of multiple memories.} Each latching pair like Fig.~\ref{fig:mech}d corresponds to a point on a plane, according to its upper thresholds. 
\textbf{(a)} An infinite ensemble of two-hysteron latches with continuously-distributed thresholds, all initialized to the $--$ state. The thresholds are labeled $\gamma_A$, $\gamma_B$ for convenience. In accordance with Fig.~\ref{fig:mech}d, $\gamma_B > \gamma_A > 0$.
\textbf{(b)} ``Template'' for how an asymmetric cycle with amplitude $\gamma_0$ changes the states of latches in the ensemble.
\textbf{(c)} The ensemble in (a), after one cycle with amplitude 4\% forms a memory.
\textbf{(d)} The ensemble encodes two memories after cycles with 4\% and 2\% amplitude.
\textbf{(e)} Readout of the ensemble in (d), reporting the fraction of hysterons in different states, in latches that conform to Fig.~\ref{fig:mech}d and/or Fig.~\ref{fig:mech}e. Curves show readout of all latches (``all''), and of ensembles generated separately to match each labeled region in (d).
}
\label{fig:plane}
\end{figure}

\subsection{Nesting Memories of Asymmetric Driving}

Finally we return to our original question: how a system with hysteresis may store multiple memories of asymmetric driving. 
We focus on the upper thresholds of the two-hysteron latch, $\gamma_1^+(-)$ and $\gamma_2^+(+)$, which we relabel $\gamma_A$ and $\gamma_B$ for convenience, with $\gamma_B > \gamma_A$. In Fig.~\ref{fig:mech}d, and in general, these thresholds determine which state a latching pair lands in at the end of a shear cycle of amplitude $\gamma_0$. There are three possibilities: the latch may be left totally undisturbed in the $--$ state ($\gamma_A > \gamma_0$), it may be stuck in the $+-$ state ($\gamma_B > \gamma_0 > \gamma_A$), or it may be pushed all the way to $++$ so that it returns to the $--$ state by the end of the cycle ($\gamma_0 > \gamma_B$).

\textbf{Disordered systems ---} An ensemble of many latches can be represented as a set of points on the $\gamma_B$-$\gamma_A$ plane, where $\gamma_B > \gamma_A > 0$, 
depicted in Fig.~\ref{fig:plane}a with all latches initialized to $--$. Figure~\ref{fig:plane}b illustrates the result of a cycle with amplitude $\gamma_0$: latches with $\gamma_A > \gamma_0$ are undisturbed, those with $\gamma_B < \gamma_0$ leave but return to the $--$ state, and all others go to the $+-$ state. 
By applying the template of Fig.~\ref{fig:plane}b repeatedly with different $\gamma_0$, we can graphically find the state of the ensemble after an arbitrary sequence of amplitudes. For example, in Fig.~\ref{fig:plane}(c, d) cycles of amplitude 4\% and then 2\% write two memories, as in the ``4, 2'' protocol of Fig.~\ref{fig:protocol}b; the memories form a ``stair-step'' pattern on the plane. 

To generate a readout signal like Fig.~\ref{fig:mech}c, we apply cycles of increasing amplitude.
The horizontal edge of the template, positioned at $\gamma_0$ in Fig.~\ref{fig:plane}b, becomes a front that starts at zero and moves upward, changing pairs to $+-$ as it passes; the vertical edge starts at zero and moves rightward, changing pairs to $--$.
In Fig.~\ref{fig:plane}e we show how each region marked in Fig.~\ref{fig:plane}d contributes to a distinctive signal.
Pairs in the closed regions ``i'' and ``iii'' 
begin readout as $--$, are changed to $+-$ by the horizontal front and add to the readout signal, and then are changed back to $--$ by the vertical front, forming non-monotonic contributions to the readout. For the pairs in regions ``ii'' and ``iv'' that were placed in the $+-$ state by writing memories, the horizontal front has no effect, but the vertical front changes them to $--$, so that their contribution to readout instead rises monotonically for all subsequent $\gamma_\text{read}$. Finally, when readout surpasses the largest stored memory (4\%) it reaches the triangular region ``v'', which extends to the largest $\gamma_A$ and $\gamma_B$ allowed in our simulation, beyond the limits of the plots. This region's contribution rises steeply as the horizontal front begins to sweep over it, falling only at much greater $\gamma_\text{read}$ (not plotted) when our finite ensemble becomes saturated. Altogether, the two stored memories create exactly two cusps in the combined readout signal, where its slope increases discontinuously. 


The equivalence in Fig.~\ref{fig:mech} between a single hysteron under symmetric driving and a frustrated pair under asymmetric driving suggests that their multiple-memory capacities may be understood in the same way. Indeed, our method can describe the return-point memory of symmetric cycles that begin with positive strain, via the change of variables $\gamma_A \to -\gamma^-$, $\gamma_B \to \gamma^+$. However, our scheme is distinct from earlier graphical analyses of return-point memory for arbitrary driving (neither symmetric nor asymmetric)~\cite{barker1983,Keim2020GlobalSolid,Keim2022MechanicalSolid,Preisach1935}. 
In the Supplemental Materials~\cite{SM} we extend our analysis to arbitrarily many memories and we consider the areas of the stair-steps of Fig.~\ref{fig:plane}d to show that, as with return-point memory, the maximum number of nested memories scales with the square root of the ensemble size.

\textbf{Designed systems and sequence recognition ---} The two-dimensional analysis in Fig.~\ref{fig:plane} can be made nearly one-dimensional if the hysterons' parameters can be specified: only some latches near the $\gamma_A = \gamma_B$ line are needed to encode the stair-step signature of nested memories, and the rest are redundant. We demonstrate this idea by constructing an ensemble. First, we divide the entire range of expected amplitude values into $M$ non-overlapping intervals: 
$[\tilde \gamma_0, \tilde \gamma_1),
[\tilde \gamma_1, \tilde \gamma_2), \dots, [\tilde \gamma_{M-1}, \tilde \gamma_M)$. 
The two endpoints of each interval then become $\gamma_A$ and $\gamma_B$ for each of $M$ latches. As in the preceding discussions, finding the $m$th latch latch in the $+-$ state is evidence that a cycle with an amplitude between $\gamma_A = \tilde \gamma_{m-1}$ and $\gamma_B = \tilde \gamma_{m}$ was applied, and that no cycle exceeding $\tilde \gamma_m$ has been applied since then. Thus $M$ latches digitize and store $M$ distinct amplitudes---a linear scaling, instead of the square-root scaling in a disordered ensemble. Together, the latches' states are bits that distinguish one sequence of nested amplitudes from among $2^M$ possibilities. 
Testing the states of these bits forms the basis for a combination lock.

\section{Discussion}

Return-point memory is a recipe for retaining arbitrarily many values from the history of a single variable, by coupling that driving to an ensemble of hysteretic elements. It has numerous examples in the natural world and in engineering, and in many cases (most importantly, when no frustrated interactions are present) it is the only possible behavior. It nonetheless fails whenever driving is asymmetric or rectified, as in a pedal depressed multiple times, or electrical signals from flashes of light.

Our results show that it is possible to store details of asymmetric driving if a system's hysteretic elements interact. The behavior clearly violates return-point memory, since the driving is bounded between two turning points yet revisiting one of those points yields a new state. Nonetheless, the similarities with return-point memory are striking. Both mechanisms always store the most recent input, but preserve past memories when amplitude is reduced, so that a system encodes the history of nested cycles of decreasing amplitude. Each kind of memory allows previous states to be recalled as amplitude is increased, yielding similar readout curves. Each arises from the smallest and simplest characteristic unit of its system: a ``latch'' formed by a single hysteron or an interacting pair. Finally, each memory behavior is dominant for its respective driving type, even if these units interact with their environments, permitting the mechanisms to be highly scalable and defect-tolerant. 

Together, these two behaviors point to a principle even more generic than return-point memory: robust, nested memories arising from units that ``latch`` at some input value and reset at a larger value, as in the parallel diagrams of Fig.~\ref{fig:mech}(b, d). For a single hysteron, this pattern is realized by the  
asymmetric placement of the flipping strains around 0, 
while for the latching pair, one hysteron cannot return to a ``down'' state until a large deformation drives another hysteron ``up''---the essence of a frustrated interaction.

Recent progress in creating and describing interacting mechanical hysterons~\cite{Jules2022Origami,bense2021complex,Sirote-Katz2024,Liu2024,Muhaxheri2024} has already led to experiments with highly tunable interactions~\cite{Liu2024}. Because the two-hysteron latching behavior is described by a simple rule (Eq.~\ref{eq:simpler_pair}), it may be implemented in such experiments as a mechanism for multiple memories or sequence recognition.


Our work also points to new opportunities for the study of glassy matter. Even though frustrated interactions are essential to the physics of amorphous solids~\cite{kumar2022,nicolas2018}, crumpled sheets~\cite{Shohat2022MemorySheets}, and some magnetic systems~\cite{binder1986,hovorka2005,gilbert2015} and mechanical metamaterials~\cite{Merrigan2021Metamaterial,Sirote-Katz2024}, in existing memory studies frustration has largely been relegated to perturbing return-point memory. 
Our results show that a simple change to the driving protocol can suppress return-point memory and reveal a rich, intelligible, and distinctly glassy form of memory. Because this memory arises from a single dominant mechanism, even in larger systems, experiments and molecular dynamics simulations can characterize interaction strengths by tracking individual relaxations while varying the amplitude and origin of asymmetric driving. More generally, the readout method is based on differences, and so we look forward to results like those in Fig.~\ref{fig:double}b that quantify frustration in macroscopic samples via measurements of magnetization, light scattering, or even image subtraction~\cite{Keim2022MechanicalSolid}.

Our study adds to the evidence that frustrated matter can remember what return-point memory must forget---that weakly breaking return-point memory tends to expand memory capacity. This hypothesis is also supported by studies of glassy systems' ability to count repeated cycles~\cite{Keim2021MultiperiodicSolids,Lindeman2021MultipleLandscapes,vanHecke2021ProfusionHysterons,Szulc2022CooperativeSolids} and retain vestiges of erased memories~\cite{Lindeman2021MultipleLandscapes}---although in those examples the mechanisms are unclear or lack a common motif, and they require $N \ge 3$ hysterons. By contrast, the two-hysteron latch is a singular mechanism that is as small as possible, yet scales linearly to store arbitrarily long sequences. The elementary principles and designs emerging from our work and from other recent studies hold promise for new kinds of mechanical information-processing systems as useful, robust, and ubiquitous as the venerable combination lock.

\begin{acknowledgments}
We thank Martin van Hecke, Mahesh Bandi, and Surendra Padamata for insightful conversations. CWL was supported by the US Department of Energy, Office of Science, Basic Energy Sciences, under Grant DE-SC0020972. TRJ was supported in part by a Remote Innovation Grant from the Student Engagement Network at Penn State, and by a grant from the Schreyer Honors College at Penn State. 
All authors wrote the paper. CWL performed theoretical analyses, TRJ and NCK performed simulations and analyzed results, and NCK designed the study.
The authors declare that they have no competing interests. 
All data needed to evaluate the conclusions in this paper are present in the paper, the Supplementary Materials, and in an archive deposited at \texttt{https://zenodo.org/XXXXXX}.

\end{acknowledgments}

\bibliography{apssamp}

\onecolumngrid

\end{document}


\renewcommand{\thefigure}{S\arabic{figure}}
\setcounter{figure}{0} 
\renewcommand{\theequation}{S\arabic{equation}}
\setcounter{equation}{0}
\renewcommand{\thetable}{S\arabic{table}}
\setcounter{table}{0}

\title{Supplementary Material for \\ ``Generalizing multiple memories from a single drive: The hysteron latch''}

\date{\today}

\maketitle

\renewcommand{\thefigure}{S\arabic{figure}}
\setcounter{figure}{0} 
\renewcommand{\theequation}{S\arabic{equation}}
\setcounter{equation}{0}





\section{Effect of initial condition}

\begin{figure}[!ht]
    \begin{center}
        \includegraphics[width=\textwidth]{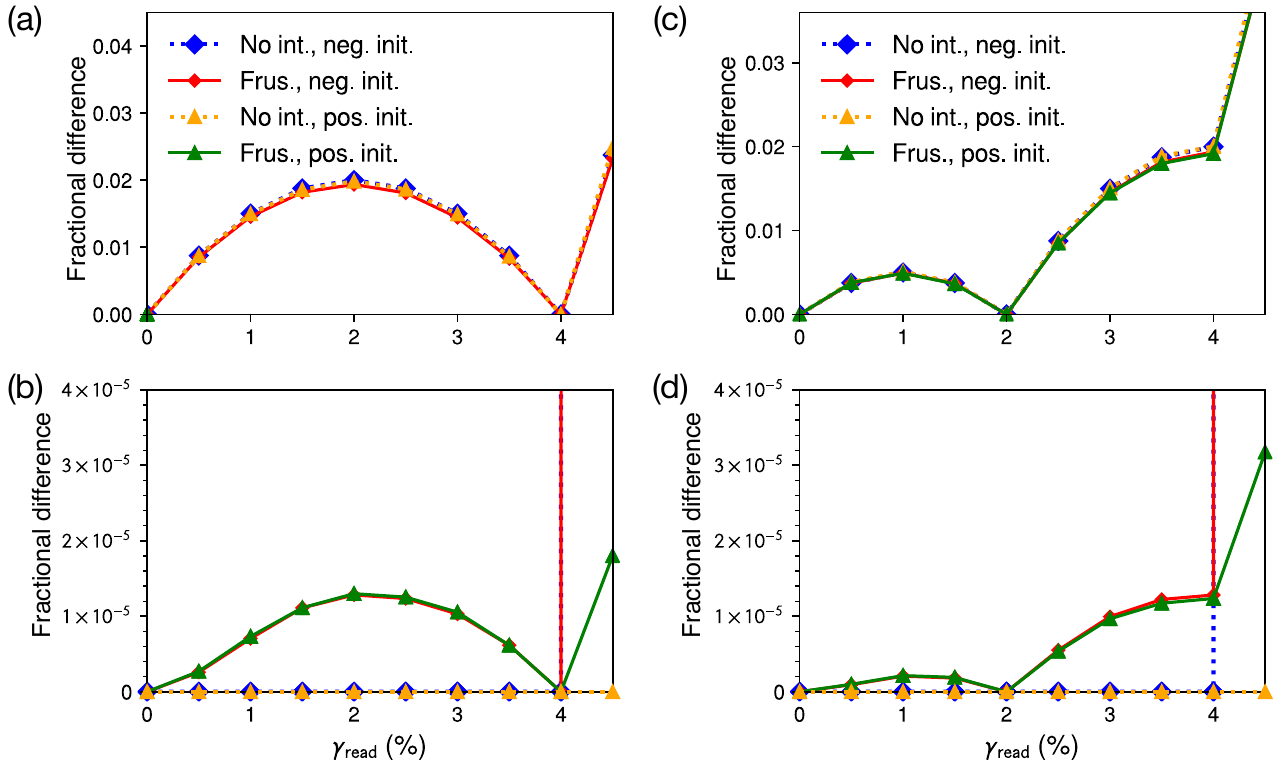}
    \end{center}
    \caption{
        \textbf{Readout curves showing role of initialization for asymmetric driving.}
        Data are reproduced from Figs.~3 and 4(a,c), with the addition of frustrated ensembles starting from an all-positive initial state (i.e.\ the same polarity as driving).
        \textbf{(a)} Readouts after writing a single memory at $\gamma = 4\%$.
        \textbf{(b)} Readouts after writing memories at 4\% and 2\%. Positive initialization has no effect on the two-hysteron mechanism that gives rise to the signal for $\gamma_\text{read} \le 4\%$, while it changes the slope of the response at larger $\gamma_\text{read}$.
        \label{figsi:init}
    }
\end{figure}

In the main text we consider asymmetric driving with positive strain only, and all hysterons beginning in the ``--'' state---opposite the polarity of driving, as shown in Fig.~1. Here we describe simulations that begin in the ``+'' state, as though $\gamma \to \infty$; in these simulations we drive the hysterons to $\gamma = 0$ and then write and read memories as in Figs.~1 and 2. 
Figure~\ref{figsi:init} shows these results along with the curves from Figs.~3 and 4(a,c). 
The effect on \emph{symmetric} driving in Fig.~\ref{figsi:init}(a, c) is subtle. However, the effect of initialization on return-point memory under \emph{asymmetric} driving in Fig.~\ref{figsi:init}(b, d) can be dramatic. 
In simulations without interactions ($J_{ij}=0$) the asymmetric curve initialized with the same (positive) sign shows no memory at all---it has a fractional difference of 0 for all readout strains. This is a consequence of return-point memory: because $\gamma$ is bounded between $\infty$ and 0, visiting the turning point at 0 after each cycle restores the system to its state before encoding. 
By contrast, the second-order memory mechanism illustrated in Fig.~4d is not return-point memory, and its behavior for $\gamma \le 4\%$ is the same whether we first approach $\gamma = 0$ from $+\infty$ or $-\infty$. Thus in Fig.~\ref{figsi:init}b, the two asymmetric protocols each form a memory, and their readout results agree for $\gamma_\text{read} < 4\%$. Differences at larger $\gamma_\text{read}$ are due to hysterons that were not affected by writing and readout, and that retain only a memory of the initial condition.

Asymmetric driving with same-polarity initialization is thus a special case for return-point memory, and for simplicity we have omitted it from the main text. Other initialization protocols~\cite{Keim2022MechanicalSolid} will generally be able to form memories of single amplitudes under asymmetric driving, even in systems with perfect return-point memory. We note that the readout behavior past the stored memory (here, at $\gamma_\text{read} = 4\%$) is itself a kind of memory of initial conditions.

\section{Interaction strength}

There are three strain scales that play a role in the effect of interactions on a pair of hysterons: the amount of hysteresis before adding interactions $\gamma^+ - \gamma^-$ (sometimes called the hysteron length), the interaction strength scale $J_0$, and the driving amplitude, taken here to be a few percent, $\gamma_0 \approx 0.04$.  The effect of the driving amplitude is to select which hysterons are involved in the writing and readout process: only hysterons with hysteresis $\gamma^+ - \gamma^- \sim \gamma_0$ can contribute. Here, the hysteresis is effectively chosen from $[0,0.2]$, and the contributing hysterons are further cut down by $\gamma_0 \approx 0.04$. A natural non-dimensional scale of interaction strength would therefore be relative to the driving amplitude: $J^* \sim J_0/\gamma_0$. In our scaling study, $J_0$ ranges from $10^{-5}$ to $10^{-2}$, giving $J^*$ from $\sim 3 \times 10^{-4}$ to $\sim 0.3$. 

Note that in other cases where hysteresis is small relative to the driving amplitude, the non-dimensional scale of interaction strength might be better defined as the ratio of $J_0$ to the hysteron length.

\section{Readout curves for various $J_0$ and $N$}

\begin{figure}[!ht]
    \begin{center}
        \includegraphics[width=\textwidth]{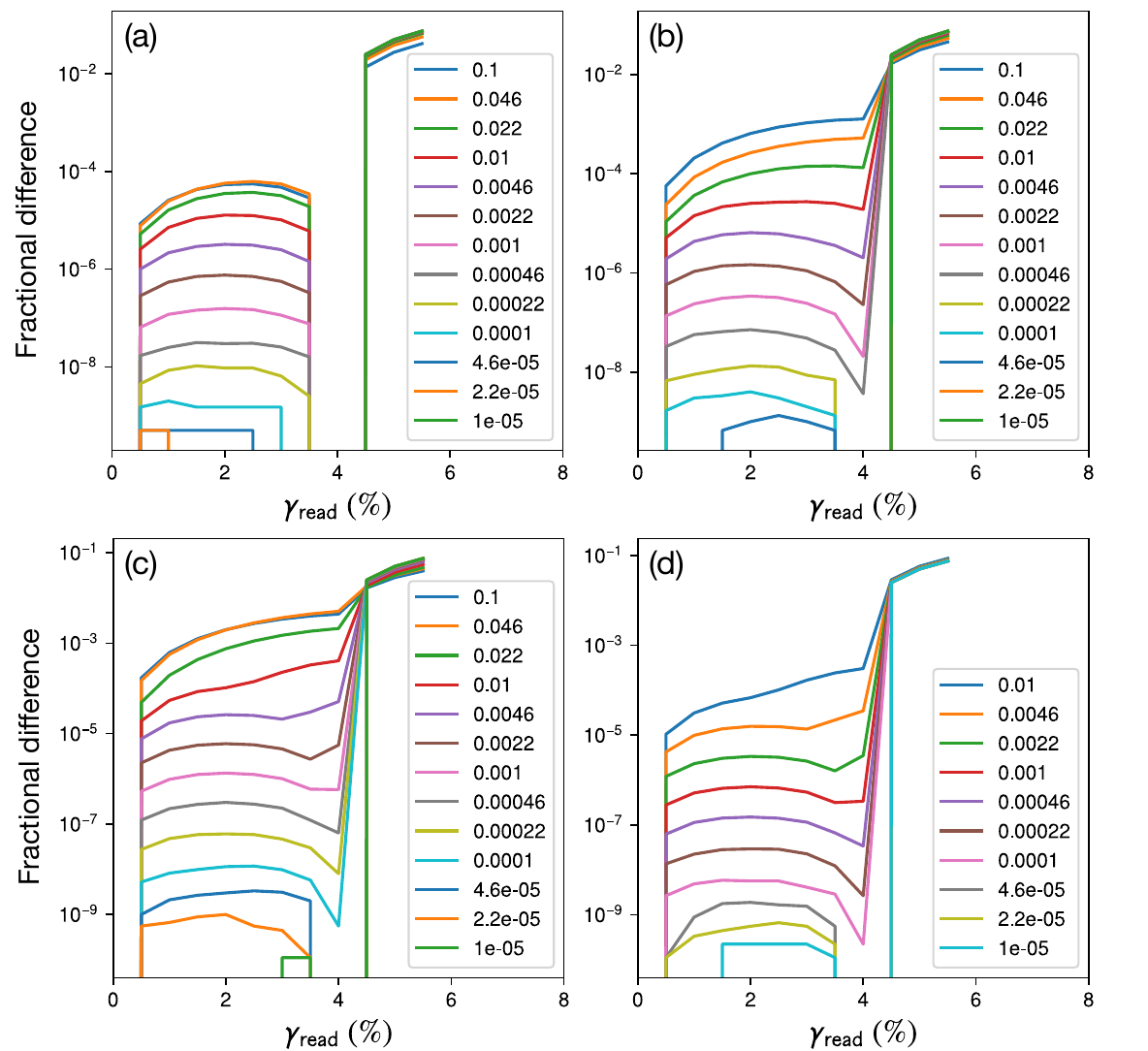}
    \end{center}
    \caption{
        \textbf{Single-memory readout curves for various interaction strength scales $J_0$ and number of mutually-interacting hysterons $N$.} Each panel corresponds to a curve in Fig.~4f: (a) $N = 2$; (b) $N = 3$; (c) $N = 9$; (d) the ``$N = 9\pm$'' case where $J_{ij}, J_{ji}$ pairs are made positive with probability 1/2. The legend identifies the value of $J_0$ for each curve and matches the vertical order of the curves; larger $J_0$ generally result in larger fractional differences. Fractional differences of zero are not shown on these logarithmic axes. Each curve was generated from an ensemble of $10^9$ groups of $N$ hysterons.
        \label{figsi:readoutsJN}
    }
\end{figure}

Figure~\ref{figsi:readoutsJN} shows the readout curves, generated in the manner of Fig.~4c, that were used to measure memory signal strength for Fig.~4f. The ``maximum memory signal'' plotted in the inset of Fig.~3b refers to the maximum value of each curve for $\gamma_\text{read} < 4\%$, and is measured only on curves that are non-monotonic, meaning that some larger values of $J_0$ are not plotted in the paper.
We also omit measurements for $N = 2$, $J_0 < 10^{-4}$, for $N=3$, $J_0 < 4.6 \times 10^{-5}$, and for $N = 9$, $J_0 < 2.2 \times 10^{-5}$ because in those cases the entire signal comes from at most 1 hysteron, making a measurement difficult to interpret.

\section{Transition graphs and generalized inequalities}

\begin{figure}
\includegraphics[width=16.5cm]{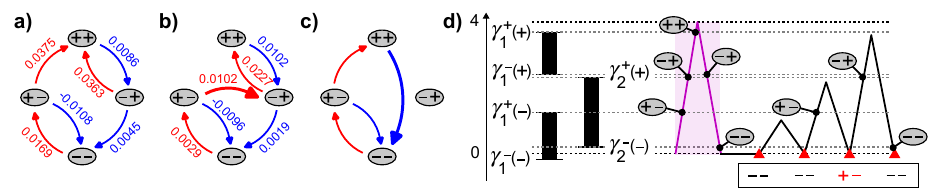}
\caption{\textbf{Two-hysteron transition graphs that give rise to non-zero readout below the remembered strain.} Transitions encountered upon increasing (decreasing) strain are in red (blue). Examples of strain values that give rise to such a behavior are shown.
(a) Transition graph corresponding to the example in Fig.~4d, also shown in Fig.~4e.
(b) A second contributing transition graph, which occurs when $\gamma_1^-(+) > \gamma_2^+(+)$, also shown in Fig.~4e. This causes a so-called ``horizontal'' avalanche when the second hysteron flips to the $+$ state under increasing strain~\cite{vanHecke2021ProfusionHysterons}.
(c) A third contributing transition graph, which occurs when $\gamma_2^-(-) > \gamma_1^-(+)$. This causes a so-called ``vertical'' avalanche when the first hysteron flips to the $-$ state under decreasing strain. No example strains are given because strictly negative interactions in the $N=2$ pairs simulated prevented such a graph from contributing to readout.
(d) Example hysterons in the strain-space format of Fig.~4(b,d) of the main text and corresponding to the type of transition graph shown in (b) (the flipping strains shown are not the exact values marked in (b); rearrangements have been spaced out for clarity).}
\label{figsi:t-graphs}
\end{figure}

A particular group of hysterons can be represented as a transition graph, in which nodes are states of the hysterons (for example, $++$ or $+-$) and arrows represent transitions between states for increasing and decreasing strain. For two hysterons, all such transition graphs are enumerated in~\cite{vanHecke2021ProfusionHysterons}. Of those, only three can contribute to the non-zero readout signal explored in Fig.~4. These transition graphs are shown in Fig.~\ref{figsi:t-graphs}, with example values of the flipping strains given for each arrow in the cases observed in simulation. These example flipping strains satisfy Eq.~2.

Note that, in practice, the transition graph in Fig.~\ref{figsi:t-graphs}b did not contribute substantially to the readout signals, as it requires large interaction strength relative to the length $\gamma_2^+ - \gamma_2^-$ of hysteron 2. We quantify this contribution in the bottom row of Table~\ref{tab:kine}. The transition graph in Fig.~\ref{figsi:t-graphs}c did not contribute at all because it requires breaking the rule used throughout the main text that each $J_{ij}, J_{ji}$ pair must have the same sign~\cite{vanHecke2021ProfusionHysterons}.

To generalize the inequalities in Eq.~2 to include all three transition graphs, we allow for non-frustrated interactions. For the latching mechanism necessary for non-zero readout to occur, we still need one frustrated interaction (without loss of generality, $J_{12} < 0$), so the only new case allowed by this relaxation of strictly frustrated interactions is $J_{21} > 0$. 

The general picture that emerges from the three contributing diagrams (shown in Fig.~\ref{figsi:t-graphs}a--c) is (1) that the hysterons must travel through all four possible states in a cycle of sufficient size (rather than, for example, going through $+-$ on both the way up and the way down); and (2) that the transition back from $+-$ to $--$ is outside the range of driving, that is, below zero. This sets up the following inequalities:
\begin{itemize}
\item If we define hysteron 1 as the hysteron that flips forward first, then hysteron 1 should also be the first to flip back: $\gamma^-_1(+) > \gamma^-_2(+)$. This makes it possible to travel through all four states in a shear cycle.
\item The transitions from $++$ to $-+$ and from $-+$ to $--$ must occur at $\gamma > 0$. In other words, since we have already determined that hysteron 1 is the first to flip back to $-$, $\gamma^-_1(+) > 0$ and $\gamma^-_2(-) > 0$.
\item The transition from $+-$ to $--$ must be below zero: $\gamma^-_1(-) < 0$. This prevents the pair from always returning to the same state when the strain is returned to 0.
\end{itemize}
More concisely, the necessary and sufficient set of inequalities to get non-zero readout below the remembered strain for two interacting hysterons that have been driven asymmetrically is 
\begin{align}
\gamma^-_1(-) &< 0 < \gamma^-_1(+) \\
\gamma^-_2(+) &< \gamma^-_1(+) \\
0 &< \gamma^-_2(-),
\label{eqsi:general_ineq}
\end{align}
which is satisfied by graphs \ref{figsi:t-graphs}(a--c) and in Figs.~4d and \ref{figsi:t-graphs}d, and of which Eq.~2 is the special case for $J_{21} < 0$.

\section{Frustrated-Pair Mechanism in Larger Systems}

\begin{table}
    \centering
\begin{tabular}{lrlrlrlrlr}
\toprule
\hfill $N$ &                   $2$ & $\ $ &                   $3$ & $\ $   &                   $9$ & $\ $    &                $9\pm$~(a) & $\ $     &         $9\pm$~(b) \\
\midrule
$P(\text{coop})$                                   &                     0 &      &                     0 &        &                     0 &         &                 $1/2$ &          &                 $1/2$ \\
$J_0$                                              & $1.00 \times 10^{-2}$ &      & $2.00 \times 10^{-3}$ &        & $2.00 \times 10^{-3}$ &         & $2.00 \times 10^{-3}$ &          & $1.00 \times 10^{-2}$ \\
$P(\text{sig})$                                    & $2.07 \times 10^{-5}$ &      & $2.19 \times 10^{-6}$ &        & $8.71 \times 10^{-6}$ &         & $4.72 \times 10^{-6}$ &          & $7.82 \times 10^{-5}$ \\
$P(\text{sig}) / (N - 1) (1 - P(\text{coop})) J_0^2$ &                $0.21$ &      &                $0.27$ &        &                $0.27$ &         &                 $0.30$ &          &                 $0.20$ \\
$P(\text{kine} \mid \text{sig})$                   &                   $1$ &      &                $0.98$ &        &                $0.91$ &         &                $0.82$ &          &                $0.67$ \\
$P(\text{param} \mid \text{kine})$                 &                   $1$ &      &                $0.24$ &        & $7.21 \times 10^{-2}$ &         & $4.57 \times 10^{-2}$ &          & $4.69 \times 10^{-2}$ \\
$P(\text{coop} \mid \text{kine})$                  &                      &      &                      &        &                      &         & $1.22 \times 10^{-2}$ &          & $6.09 \times 10^{-2}$ \\
$\langle J_{12} \rangle / J_0$                     &               $-0.78$ &      &                $-0.80$ &        &                $-0.80$ &         &               $-0.79$ &          &               $-0.77$ \\
$\langle J_{21} \rangle / J_0$                     &               $-0.18$ &      &                $-0.20$ &        &                $-0.20$ &         &               $-0.21$ &          &                $-0.20$ \\
$P(\text{avalanche} \mid \text{kine})$             & $2.24 \times 10^{-2}$ &      &                     0 &        &                     0 &         &                     0 &          & $2.20 \times 10^{-2}$ \\
\bottomrule
\end{tabular}

    \caption{\textbf{The pair mechanism of Fig.~4d dominates memory of asymmetric driving in larger groups of hysterons.} Analysis was performed on ensembles of $10^9$ groups of size $N$ (columns); for the two ``$9\pm$'' ensembles, pairs of $J_{ij}, J_{ji}$ were made positive with probability $1/2$. Analysis identified hysterons that contributed to the readout signal (``sig''), that matched the kinematics of Figs.~4d and/or \ref{figsi:t-graphs}d (``kine''), and that had cooperative interactions (``coop'') or horizontal avalanches as in Fig.~\ref{figsi:t-graphs}d (``avalanche''). $P(\text{sig})$ is also rescaled according to Fig.~4f. Average interaction values $\langle J_{12} \rangle, \langle J_{21} \rangle$ are from hysteron pairs that match kinematics and have a frustrated interaction; subscripts refer to the corresponding hysteron indices in Fig.~4d. $\langle J_{12} \rangle, \langle J_{21} \rangle$ are remarkably consistent, suggesting that memory kinematics of a pair are driven primarily by their interaction, even in larger groups.}
    \label{tab:kine}
\end{table}

The scaling analysis in Fig.~4f suggested that the memory of asymmetric driving in larger groups of $N$ mutually-interacting hysterons (Fig.~\ref{figsi:readoutsJN}) was dominated by the action of frustrated pairs. While for $N=2$ one can predict memory behavior by inspecting the model parameters (Eqs.~1 and 2), for $N > 2$ the numerous interactions and metastable states make this impractical. Instead, we can investigate the origins of memory readout signals in larger groups and ask whether the mechanism for memory in a pair is at work, by examining kinematics: the transitions of individual hysterons within larger groups, as a memory is written and read with the asymmetric protocol in Fig.~2c. 

Table~\ref{tab:kine} summarizes our analysis of the kinematics in ensembles of $10^9$ groups, generated according to the same schemes as the inset of Fig.~3b and Fig.~\ref{figsi:readoutsJN}. We fix the maximum interaction strength at $J_0 = 2\times 10^{-3}$, except $J_0 = 10^{-2}$ for $N=2$, the $N$ for which the memory behavior is least likely in a group.
We first drive an ensemble with the single-memory asymmetric protocol in Fig.~2a. We identify the probability $P(\text{sig})$ that a given hysteron contributes to a non-monotonic memory readout signal, as in Fig.~4c --- specifically, that its state after the cycle with $\gamma_\text{read} = 4\%$ matches its state at the beginning of readout, and that it is in a different state at the end of at least one intervening cycle.

\begin{figure}
\includegraphics[width=0.9\textwidth]{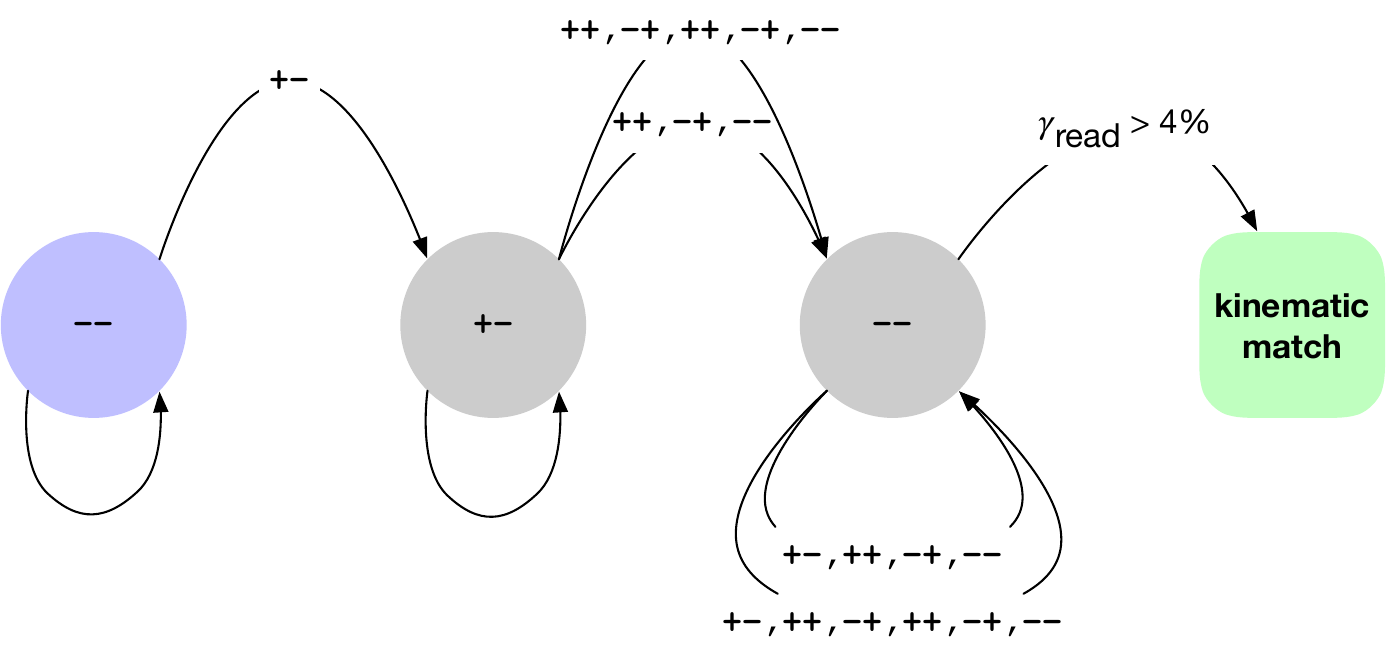}
\caption{\textbf{Testing the kinematics of a hysteron pair.}
A finite-state machine tests whether the kinematics of a memory-encoding pair during readout match the readout sequence shown in Fig.~4d or Fig.~\ref{figsi:t-graphs}d. Each arrow represents one cycle of asymmetric driving and shows the sequence of states (if any) during that cycle; each node represents the state at the beginning/end of the cycle. Readout begins in the $--$ state at left and proceeds with increasing $\gamma_\text{read}$. To be accepted as a kinematics match, a hysteron pair must follow only an allowed sequence of states, and must be in the right-hand $--$ state after the cycle with $\gamma_\text{read} = 4\%$ (the strain amplitude that was written). The longer sequences from $+-$ to $--$ and from $--$ to itself correspond to the uppermost transitions in Fig.~\ref{figsi:t-graphs}b.
}
\label{figsi:kinematics}
\end{figure}

Next, each hysteron that counts toward $P(\text{sig})$ belongs to several interacting pairs (for $N$ hysterons, $N-1$ pairs). For each pair, we extract the sequence of states that the pair visits in each cycle of readout. Figure~4b establishes that the memory behavior of interest arises from the distinct sequences obtained during small-, medium-, and large- amplitude cycles. To check whether these sequences match the progressions in Fig.~4d or Fig.~\ref{figsi:t-graphs}d, our analysis implements the finite-state machine in Fig.~\ref{figsi:kinematics}; this also checks for a (rare) variant made possible by the avalanche behavior. We then ask: of the hysterons that contribute to the readout signal, what fraction $P(\text{kine} \mid \text{sig})$ belongs to exactly one pair that matches these kinematics? Remarkably, in Table~\ref{tab:kine} we report that $P(\text{kine} \mid \text{sig}) \sim 1$ even in large groups with cooperative and frustrated interactions (``$9\pm$'').

We can further ask whether the actual mechanism of Fig.~4d, which is based on an isolated pair of hysterons, has any explanatory value for the pairs embedded in larger groups. At one extreme, one might hypothesize that these pairs are nearly isolated from their companions, so that their randomly-generated interaction \emph{and} threshold parameters are often consistent with Eq.~2---a probability we measure as $P(\text{param} \mid \text{kine})$. Instead, Table~\ref{tab:kine} shows that $P(\text{param} \mid \text{kine}) \ll 1$ in large groups, meaning that the effects of other hysterons on the thresholds in Eq.~2 are significant; the behavior cannot be predicted from the two hysterons alone. 

At the other extreme, one might hypothesize that a pair of hysterons with memory kinematics is merely a ``puppet'' of other hysterons with unknown memory behaviors, and that the interaction within the pair is so unimportant that it need not even be frustrated. In the ``$9\pm$'' ensembles where cooperative interactions occur with probability 0.5, we find that this ``puppet'' scenario is not supported, since $P(\text{coop} \mid \text{kine}) \ll 1$. 

Most remarkably, Table~\ref{tab:kine} shows that for hysteron pairs that match kinematics and have a frustrated interaction, the mean interaction values $\langle J_{12} \rangle, \langle J_{21} \rangle$ are nearly the same fraction of $J_0$ in all ensembles. In other words, the kinematics of a memory-forming pair within a larger group are driven by the interactions within that pair, in the same way as for $N=2$. 
Together, the results in Table~\ref{tab:kine} strongly support our interpretation of the $J_0^2$ scaling results in Fig.~4f: the mechanism in Fig.~4d is the dominant way that hysterons remember asymmetric driving, even in large systems with a mixture of cooperative and frustrated interactions.  



\section{Generalizing multiple-memory capacity in disordered ensembles}

\begin{figure}
\includegraphics[width=\textwidth]{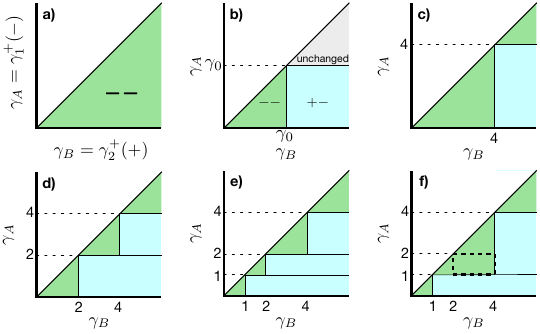}
\caption{\textbf{Graphical analysis of multiple memories.} Each pair of hysterons like Fig.~4d can be represented as a point on a plane, according to its upper thresholds. 
(a) An infinite ensemble of such pairs with continuously-distributed thresholds, all initialized to the $--$ state. The thresholds are labeled $\gamma_A$, $\gamma_B$ for convenience. In accordance with Fig.~4d, $0 < \gamma_A < \gamma_B$.
(b) ``Template'' for how an asymmetric cycle with amplitude $\gamma_0$ changes the states of pairs in the ensemble.
(c) The ensemble in (a), after one cycle with amplitude 4\% forms a memory.
(d) The ensemble encodes two memories after cycles with 4\% and 2\% amplitude.
(e) Ensemble after cycles with 4\%, 2\%, and 1\%.
(f) Ensemble after cycles with 4\% and 1\%. Comparing with (e) reveals the set of pairs (outlined with heavy dashes) that encode the difference between these histories.
\label{figsi:plane} }
\end{figure}



Here we generalize the template method outlined in Fig.~5 of the main text beyond the one- and two-amplitude writing sequences shown there. Figures~\ref{figsi:plane}(a--d) reproduce the initial case, the general template, and the one- and two-amplitude cases from Fig.~5(a--d). Figures~\ref{figsi:plane}(e,f) show the extension of these cases: it is possible to distinguish histories even when a larger number of amplitudes are written.

Following the argument of Keim and Medina~\cite{Keim2022MechanicalSolid}, we consider briefly the memory capacity of a finite, disordered ensemble of frustrated pairs. Figure~\ref{figsi:plane}e adds a third memory, and illustrates that nested memories form a stair-step pattern on the plot. If the density of hysteron pairs on the plane is finite, then these stair-steps can be subdivided only so many times before they become indistinct. More precisely, if we generalize our memory values as the descending sequence $\gamma_n, \gamma_{n-1}, \dots, \gamma_2, \gamma_1$, then memory $0 < m < n$ from within the sequence will be encoded by the pairs in the rectangle $\gamma_m < \gamma_B \leq \gamma_{m+1}$, $\gamma_{m-1} < \gamma_A \leq \gamma_m$. This can be seen by comparing Fig.~\ref{figsi:plane}e with the outcome if the 2\% memory had not been written (Fig.~\ref{figsi:plane}f); the highlighted rectangle encodes this difference. If there is no pair of hysterons within such a rectangle, then the ensemble’s state will be the same whether $\gamma_m$ was written or not---the memory capacity has been exceeded. Thus, in a disordered ensemble of $N$ latching pairs, the maximum number of distinct memories $M$ scales as the maximum number of distinct rectangles that can be drawn in this way, such that each rectangle contains at least one pair. The linear spacing of these finely-drawn rectangles then scales as $1/M$, and the area of each one scales as $1/M^2$.

In the case of a disordered ensemble with roughly continuous distributions of $\gamma_A$ and $\gamma_B$, the area of the smallest rectangle that contains one latching pair scales inversely with the number density of the pairs, leading to $1/M^2 \sim N$, for a memory capacity $M \sim \sqrt{N}$ as in the Preisach model~\cite{Regev2021TopologyTransition,Keim2022MechanicalSolid}. The presence or absence of each memory corresponds to a different history, and so readout lets one distinguish among $\sim 2^M$ possible histories of asymmetric driving. 

\bibliography{apssamp}